\documentclass[sigconf]{acmart}
\renewcommand\footnotetextcopyrightpermission[1]{} % removes footnote with conference information in first column
\AtBeginDocument{%
  \providecommand\BibTeX{{%
    \normalfont B\kern-0.5em{\scshape i\kern-0.25em b}\kern-0.8em\TeX}}}

\usepackage[caption=false,font=footnotesize]{subfig}
\usepackage{multirow}
\usepackage{hyperref}
\begin{document}

\title{Energy Predictive Models for Convolutional Neural Networks on Mobile Platforms}

\author{Crefeda Faviola Rodrigues}
\affiliation{%
  \institution{The University of Manchester}
  %\streetaddress{1 Th{\o}rv{\"a}ld Circle}
  %\city{Hekla}
  %\country{UK}
  }
\email{crefeda.rodrigues@manchester.ac.uk}

\author{Graham Riley}
\affiliation{%
  \institution{The University of Manchester}
  %\city{Rocquencourt}
  %\country{UK}
}
\email{graham.riley@manchester.ac.uk}

\author{Mikel Luj\'an}
\affiliation{%
  \institution{The University of Manchester}
  %\country{UK}
}
\email{Mikel.Lujan@manchester.ac.uk}

%\affiliation{%
%  \institution{Inria Paris-Rocquencourt}


\begin{abstract}
 Energy use is a key concern when deploying deep learning models on mobile and embedded platforms. Current studies develop energy predictive models based on application-level features to provide researchers a way to \textit{estimate} the energy consumption
 of their deep learning models. This information is useful for building resource-aware models that can make efficient use of the hardware resources. However, previous works on predictive modelling provide little insight into the trade-offs involved in the choice of features on the final predictive model accuracy and model complexity. To address this issue, we provide a comprehensive analysis of building regression-based predictive models for deep learning on mobile devices, based on empirical measurements gathered from the SyNERGY framework.

 Our predictive modelling strategy is based on two types of predictive models used in the literature: \textit{individual layers} and \textit{layer-type}. Our analysis of predictive models show that simple layer-type features  achieve a model complexity  of 4 to 32 times less for convolutional layer predictions for a similar accuracy compared to predictive models using more complex features adopted by previous approaches. To obtain an overall energy estimate of the inference phase, we build layer-type predictive models for the fully-connected and pooling layers using 12 representative Convolutional Neural Networks (ConvNets) on the Jetson TX1 and the Snapdragon 820 using software backends such as OpenBLAS, Eigen and CuDNN. We obtain an accuracy between 76\% to 85\% and a model complexity of 1 for the overall energy prediction of the test ConvNets across  different hardware-software combinations.  
\end{abstract}

%\begin{CCSXML}
%<ccs2012>
%<concept>
%<concept_id>10010520.10010553.10010562</concept_id>
%<concept_desc>Computer systems organization~Embedded systems</concept_desc>
%<concept_significance>500</concept_significance>
%</concept>
%<concept>
%<concept_id>10010583.10010662.10010674</concept_id>
%<concept_desc>Hardware~Power estimation</concept_desc>
%<concept_significance>500</concept_significance>
%</concept>
%<concept>
%<concept_id>10010147.10010178.10010224</concept_id>
%<concept_desc>Computing methodologies~Computer vision</concept_desc>
%<concept_significance>300</concept_significance>
%</concept>
%</ccs2012>
%\end{CCSXML}

%\ccsdesc[500]{Computer systems organization~Embedded systems}
%\ccsdesc[500]{Hardware~Power estimation and optimization}
%\ccsdesc[300]{Computing methodologies~Computer vision}

\keywords{Convolutional Neural Networks, Energy Prediction, mobile platforms}

%\keywords{datasets, neural networks, gaze detection, text tagging}


\maketitle
\pagestyle{empty}
\section{Introduction}
\label{sec:intro}

There is growing importance to bringing deep neural network processing to mobile or embedded devices (also known as \textit{edge-devices}) \citep{surveyefficientdnns}.
Deep neural networks such as Convolutional Neural Networks (hereafter referred to as ConvNets) have achieved greater accuracy compared to humans for a large variety of predictive tasks, for example, image classification in computer vision and text classification in natural language processing \citep{resnet}. %These ConvNet models are organized into \textit{layers} which are characterized by the computation they perform to transform input data into meaningful output probabilities.  
ConvNets are designed during a \textit{training phase} where machine learning researchers search for the best model using accuracy\footnote{suitably defined for the task.} as a metric. Once the ConvNet model is trained, the pre-trained model is available for use during the inference or \textit{testing} phase.

There are numerous benefits to performing inferences locally on edge devices such as reducing energy costs of datacenters, lower (user) latency and reduced need for constant internet connectivity \citep{lane}. However, such devices have a unique set of constraints in terms of resources, for example, battery life, that are atypical of the environments in which the models are trained. This has paved way for the exploration of energy-efficient ConvNet designs through manual and automated searches for low-cost neural network model designs - for example, MobileNet \citep{mobilenets} and MnasNet \citep{tan2018mnasnet} - exploration of compression and quantization and other software-based acceleration techniques, and the use of application-specific hardware accelerators \citep{surveyefficientdnns}.  \textit{Despite these efforts, there are very few studies that model the energy use of deep learning models in the context of these optimizations.}
Such modelling approaches are useful in the areas  of resource-aware ConvNet designs such as automatic \textit{Neural Architecture Search} \citep{tan2018mnasnet}, energy-aware pruning techniques \citep{yang2016designing} and in neural network accelerators simulators \citep{samajdar2018scale} that focus on designing energy-efficient deep learning hardware.

Previous studies \citep{synergy} on predictive modelling have indicated that relatively simple features, such as the sum of the multiply-accumulate (MAC) counts (we refer to this as \textit{layer-type}), can be used to estimate hardware performance counters such as SIMD instruction counts and bus accesses that are useful for determining an application's performance. The performance counter information is then used to estimate the energy consumption for the convolutional layers in a ConvNet for real systems. Other works, such as \citep{neuralpower} have relied on a  large set of complex features, extracted from each layer's specification (we refer to this as \textit{individual layers}),  to yield highly complex predictive models. However, none of these works have investigated the trade-offs of choosing features on predictive model accuracy and complexity.

The aim of this work is to perform a thorough analysis of algorithmic features of predictive models  based on layer-type and individual layers that can offer the best trade-off in model complexity (defined in Section~\ref{sec:feature-selection}) and predictive accuracy, and compare our results to previous works. %Contrary to previous works, our results show that predictive models based on simple linear layer-type features outperform complex individual layer feature-based models. These linear features can be obtained rapidly from the algorithm to provide sufficiently accurate and interpretable predictive accuracy. 
We first illustrate the techniques for layer-type versus individual layer features to build predictive models for the convolutional layers on a mobile CPU and based on the results we apply the method to other layers  in a ConvNet to get an overall estimate of the ConvNet's inference phase running on different software and mobile hardware platforms.

%First, we perform empirical power measurements on the mobile device to capture the energy use of inference workloads.  Second, we apply standard feature selection techniques to exhaustively search for relevant features to build predictive models and, third, we build energy-predictive models using regression techniques evaluating the selected features from the previous step. These predictive models are built with the view that when a new ConvNet is designed, the energy-use can be estimated rapidly and with sufficient accuracy for a given software and hardware platform, thus alleviating the need for actual energy measurements in the same environment.
\begin{figure}[t]
\centering
\includegraphics[width=0.9\columnwidth]{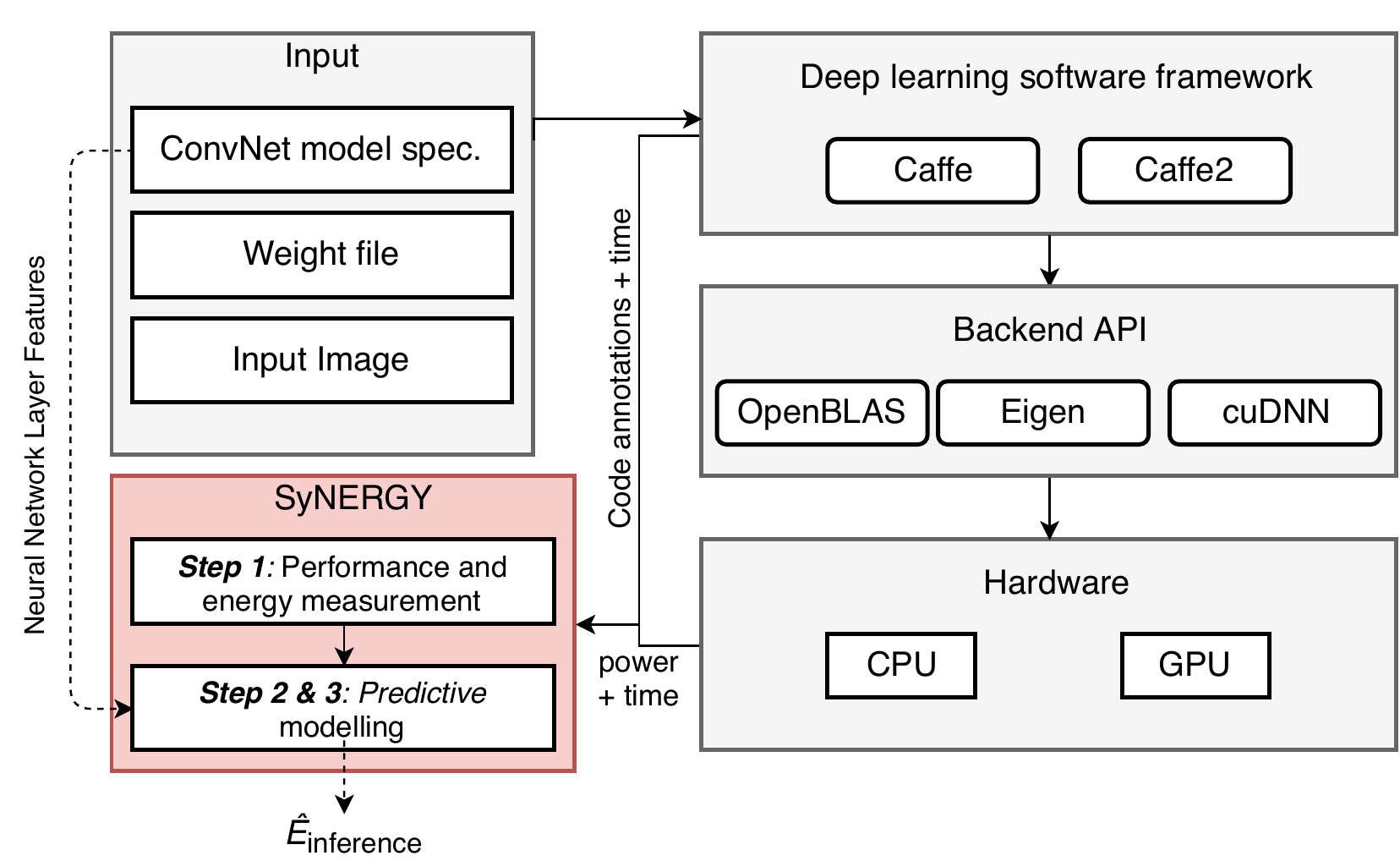}
\caption{SyNERGY energy measurement and prediction framework.}
\label{synergy}
\end{figure}

Our contribution is as follows:
\begin{itemize}
    \item An extension to the  SyNERGY framework \citep{synergy} to tie the energy measurements obtained on the mobile device to the application-level at a per layer granularity and support the building of layer-wise energy predictive models (\textbf{Step 1} in \autoref{synergy}).
    
    \item In \textbf{Step 2}, we perform an exhaustive search based on standard feature subset selection techniques to evaluate the individual layer features in terms two metrics: \textit{predictive model accuracy} and \textit{complexity} for building predictive models for the convolutional layers in the context of a given hardware-software combination. 
    We then compare the best predictive model using individual layer features with predictive models based on layer-type features to determine the best features for building predictive models in terms of predictive model accuracy and complexity.
    \item In \textbf{Step 3}, energy predictive models for different layers are built using the best features selected in Step 2. Our predictive models, based on simple algorithmic features such as the summation of multiply-accumulate (MAC) operation counts of all layers, outperform the predictive models using more complex features from the individual layers themselves. We can achieve significant accuracy with 4 to 32 times lower model complexity with similar accuracy compared to complex predictive models proposed in previous works. Our results are based on 12 representative ConvNet models chosen from existing deep learning frameworks (Caffe2 \citep{caffe}), which is a larger number than used in previous studies \citep{synergy,neuralpower} including newer low-cost ConvNet models (for example, SqueezeNet and MobileNet) used in the mobile and embedded space.
    \item We combine predictions from predictive models for different layers to get an overall estimate of the energy consumption of the deep learning model during the inference phase for multiple combinations of hardware and numerical software libraries: Eigen\footnote{\url{http://eigen.tuxfamily.org}} on a Snapdraon820 (Eigen-Snapdragon820)\footnote{\url{https://developer.qualcomm.com/software/snapdragon-neural-processing-engine}}, Eigen on a Jetson TX1 (Eigen-TX1)\footnote{\url{http://www.nvidia.com/object/jetson-tx1-module.html}} and OpenBLAS on a Jetson TX1 (OpenBLAS-TX1)\footnote{\url{http://www.openblas.net/}}. Our choice includes two platforms with the same library, that is  Eigen-Snapdragon820 and Eigen-TX1, and two libraries on a single platform, that is Eigen-TX1 and OpenBLAS-TX1. We also evaluate our methodology on a mobile GPU using CuDNN on the Jetson TX1 (CuDNN-TX1). For all hardware and software combinations, we achieve a significant predictive test accuracy in the range 76\% to 84\% compared with empirical measurements on the platforms.

\end{itemize}

The organisation of the paper is as follows. 
Section~\ref{sec:background} provides details of the ConvNet model specifications for their different layers.
Section~\ref{sec:method} goes into more detail of the specific methodology for energy measurements across the two hardware platforms: Jetson TX1 and Snapdragon 820. Section~\ref{sec:energymeasurements} presents the empirical energy measurements obtained for the overall and at different layer-types energy found in ConvNets. Section~\ref{sec:feature-selection} covers an analysis of the features to be used for the predictive models and evaluates the models based on predictive accuracy and model complexity. Section~\ref{sec:overall-energy-prediction} details the final results for the energy-predictive models for the convolutional, pooling and fully-connected layers. Section~\ref{sec:relatedwork} compares related work in performance and energy measurement and modelling. Finally, section~\ref{sec:conclusion} concludes and highlights possible future directions.
\section{ConvNet Model Specifications}
\label{sec:background}

This section covers the necessary background to understand the different layers of a ConvNet at the algorithmic level. For each layer, we describe the candidate input features (highlighted in bold) that will be evaluated in the feature selection phase in Section 5.

A ConvNet is an end-to-end pipeline of \textit{feature extraction} and \textit{classification}. The feature extractors are arranged into layers that extract high-level \textit{representations} from image data \citep{goodfellow}. As the number of layers increases the level of abstraction increases, for example, from edges or colour blobs to object shapes. The final layer uses this information to provide a classification output (or a decision). Typical layers found are convolution (Conv), pooling (Pool), normalization (there are two types: batch normalization or Local Response Normalization-(LRN)), Rectified Linear Unit (ReLU) and fully-connected (Fc) that transform the input data into a probabilistic output. We provide a description of the main layers targeted in our study: Conv, Pool and Fc.

\begin{figure}[t]
\centering
\includegraphics[width=\columnwidth]{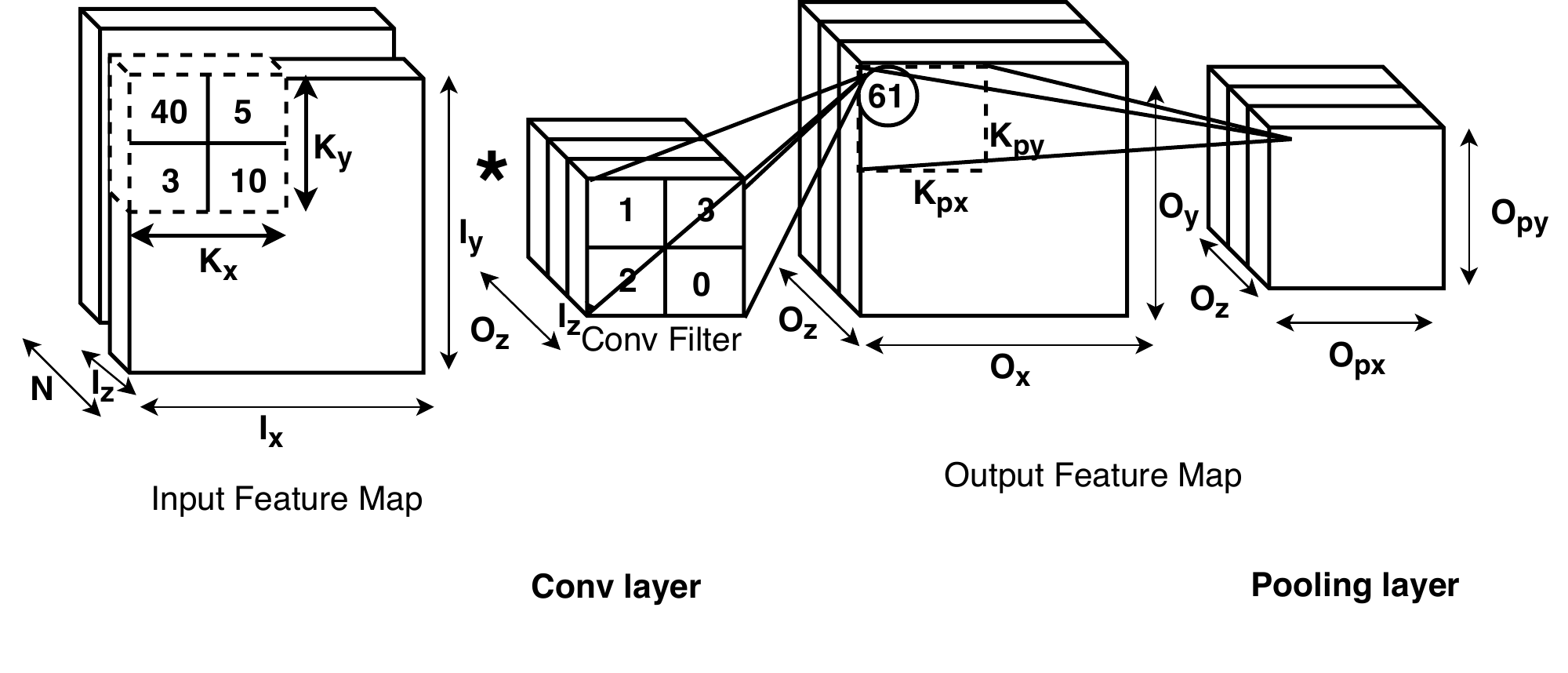}
\caption{A simple two layer ConvNet model}
\label{convnet}
\end{figure}
\begin{table*}[t]
\centering
\caption{ConvNet models in the literature}
\label{model}
\resizebox{0.9\textwidth}{!}{%
\begin{tabular}{|l|l|l|l|l|l|l|}
\hline
\textbf{ConvNet} & \textbf{\begin{tabular}[c]{@{}l@{}}Naming\\   Convention in graphs\end{tabular}} & \textbf{\begin{tabular}[c]{@{}l@{}}Top-5\\   accuracy (\%)\end{tabular}} & \textbf{Dataset} & \textbf{\# Layers} & \textbf{Parameters} & \textbf{Model Size} \\ \hline
SqueezeNet & squeezenet {\citep{squeezenet}} & 80.3 & ImageNet & 26 Conv + 3 MaxPool & 1.2 M & 5 MB \\ \hline
\begin{tabular}[c]{@{}l@{}}SqueezeNet with \\ Residual Connections\end{tabular} & squeezenetRes {\citep{squeezenet}} & 82.5 & ImageNet & 26 Conv + 3 MaxPool & 1.2 M & 6.3 MB \\ \hline
ALL-CNN-C & allcnn {\citep{allcnn}} & 90.92 & CIFAR 10 & 9 Conv & 1.3 M & 5.5 MB \\ \hline
GoogleNet & googlenet {\citep{szegedy}} & 90.85 & ImageNet & 57 Conv + 1 Fc + 13 MaxPool & 6.9 M & 54 MB \\ \hline
DenseNet & densenet {\citep{huang2017densely}} & 92.12 & ImageNet & 121 Conv + 1 MaxPool & 7.8 M & 32.3 MB  \\ \hline
Inception-v3 & inceptionv3 {\citep{szegedy}} & 90.92 & ImageNet & 94 Conv + 1 Fc + 5 MaxPool & 23 M & 95.5 MB \\ \hline
\begin{tabular}[c]{@{}l@{}}Residual-net 50 layers\end{tabular} & resnet50 {\citep{resnet}} & 93.29 & ImageNet & 53 Conv + 1 Fc + 1 MaxPool & 25 M & 103 MB \\ \hline
MobileNet & mobilenet {\citep{mobilenets}} & 70.6 & ImageNet & 27 Conv & 29 M & 17 MB \\ \hline
Places-CDNS-8s & places {\citep{placescdns}} & 86.8 & ImageNet & 8 Conv + 3 Fc + 5 MaxPool & 60 M & 241.6 MB \\ \hline
AlexNet & alexnet {\citep{imageNet}} & 80.3 & ImageNet & 5 Conv + 3 Fc + 3 MaxPool & 62 M & 244 MB \\ \hline
VGG\_CNN\_S & vggsmall {\citep{vgg}} & 86.9 & ImageNet & 5 Conv + 3 Fc + 3 MaxPool & 102 M & 393 MB \\ \hline
Inception-BN & inceptionbn {\citep{inceptbn}} & 89.0 & ImageNet & 69 Conv + 1 Fc + 5 MaxPool & 1.4 B & 134.6 MB \\ \hline
\end{tabular}
}
\end{table*}

The bulk of a ConvNet model is made up of Conv layers. The computational complexity of a standard Conv layer can be represented by the number of \textbf{Multiply-accumulates (MAC)} performed which is given by \autoref{compeq1}:
\begin{equation}
\label{compeq1}
    O_{x} \times O_{y} \times O_{z} \times K_{x} \times K_{y} \times I_{z}
\end{equation}
where $O_x$, $O_y$ and $O_z$ represent the output feature map dimensions, $K_x$, $K_y$ are the kernel filter dimensions and $I_x$, $I_y$ and $I_z$ are the input feature map dimensions in the $x$, $y$ and $z$ dimensions, as shown in \autoref{convnet}. The z dimension represents the number of channels in the feature maps.
These dimensions are governed by the \textbf{stride} (which governs the step size by which the kernel filter slides across the input in $x$ and $y$) and \textbf{padding} (the number of zeros that need to be padded around the input border to allow whole filters to be applied), and \textbf{kernel shape}.
The storage complexity or \textbf{data volume}\footnote{Referred to as Bandwidth in NeuralPower, \citep{neuralpower}} includes the cost for storing the input feature map or \textbf{input volume} to each layer, the corresponding kernel or filter \textbf{weights} ($ K_{x} \times K_{y} \times I_{z} \times O_{z}$) and biases, and the output feature map or \textbf{output volume}. The volume of data (in number) is given by \autoref{compeq2}.
\begin{equation}
\label{compeq2}
     (I_{x} \times I_{y} \times I_{z}) + (K_{x} \times K_{y} \times I_{z} \times O_{z}) + (O_{x} \times O_{y} \times O_{z})
\end{equation}
In \autoref{convnet}, $N$ refers to the number of images in the input, commonly known as the batch size.
 Newer models such as MobileNet \citep{mobilenets} leverage depth-wise separable convolutions and have the following computational complexity:
\begin{equation}
\label{compeq3}
(O_{x} \times O_{y} \times K_{x} \times K_{y} \times I_{z}) + (I_{z} \times O_{z} \times O_{x} \times O_{y})
\end{equation}

A pooling and sub-sampling layer aggregates the output from the previous layer using a pooling window ($K_{px} \times K_{py}$) in $x$ and $y$. The \textit{Max pooling} operator computes the maximum over this window and downsamples the output using the max value while the \textit{Average pooling} finds the average value over the window and down samples the output using the average value. This results in a pooling output of dimension ($O_{px} \times O_{py}$). The computation for a max value within a single window involves a comparison operation with each of its elements. For example, the $K_{px}=3$ and $K_{py}=3$ window has 9 elements that require 8 comparison operations. We refer to this as the \textbf{Op count} for the pooling layer.
\begin{equation}
\label{compeq4}
O_{px} \times O_{py} \times O_{z} \times ( K_{px} \times K_{py} -1 )
\end{equation}
    
Unlike the Conv layer, the inputs to a Fc layer are connected to all the outputs of the previous layer.  The Fc layers are similar to Conv layers with the exception that $I_{x}$, $I_{y}$, $K_{x}$, $K_{y}$ are usually greater than 1 for the first Fc layer, and then it flattens out in later Fc layers to a 1-dimensional vector.

\autoref{model} gives a list of ConvNet models chosen for this study. Column 5 represents the counts of each of the described layers - Conv, Fc and Maxpool - present in each model. Recently, in newer ConvNet models (for example, Inception-BN, Inception-V3, GoogleNet and Residual Nets) the traditional stack of fully-connected layers, seen in older models such as AlexNet and VGG, is replaced with a Global Average Pooling layer introduced in \citet*{nin} and a single Fc layer. Column 7 shows the size of the model which is stored in 32-bit floating point precision. We evaluate models ranging from 1.2M to 1.4B parameters (note, these are also referred to as weights), as given in Column 6 of \autoref{model}. The top-5 accuracy of the model is a measure based on the Top-5 predictions of the object category in a given image \citep{ImageNetChallenge}.
In our study, we evaluate the energy use of these \textit{pre-trained} ConvNets and build layer-wise energy predictive models for inferences executing on a mobile device.

\begin{table*}[t]
\centering
\caption{Platform and software specification}
\label{platform}
\resizebox{0.9\textwidth}{!}{%
\begin{tabular}{|l|l|l|l|l|l|}
\hline
\textbf{System} & \textbf{Operating System} & \textbf{\begin{tabular}[c]{@{}l@{}}Deep learning\\ framework\end{tabular}} & \textbf{\begin{tabular}[c]{@{}l@{}}Backend \\ acceleration library\end{tabular}} & \textbf{Processor} &  \textbf{Memory} \\ \hline
\multirow{2}{*}{Jetson TX1} & \multirow{2}{*}{\begin{tabular}[c]{@{}l@{}}Ubuntu 16.04 LTS\\ Linux Kernel: 4.4.38+\end{tabular}} & Caffe2 & Eigen & \multirow{2}{*}{\begin{tabular}[c]{@{}l@{}}ARM Cortex A57/A53,\\  Quad-Core,\\ 64-bit,  1.9GHz\end{tabular}} & \multirow{2}{*}{\begin{tabular}[c]{@{}l@{}}4 GB 64 bit \\ LPDDR3\\ 25.6 GB/s\end{tabular}} \\ \cline{3-4}
 &  & Caffe  & \begin{tabular}[c]{@{}l@{}}OpenBLAS, CuDNN - 6.0.21 \\ libopenblas\_-\\ cortexa57p-r0.3.1.dev.so\end{tabular} &  &  \\ \hline
\begin{tabular}[c]{@{}l@{}}Open-Q 820 \\ (APQ8096) (Intrinsyc)\end{tabular} & \begin{tabular}[c]{@{}l@{}}Android 7.0,  API: 24.0\\  (8096\_Open-\\ Q\_820\_Android\_BSP -N-3.3)\end{tabular} & Caffe2 & Eigen & \begin{tabular}[c]{@{}l@{}}Qualcomm Kryo CPU, \\ Quad-Core, \\  64-bit,  2.2GHz\end{tabular} & \begin{tabular}[c]{@{}l@{}} 3 GB 2 x 32 bit\\  LPDDR4\\ 29.9GB/s\end{tabular} \\ \hline
\end{tabular}}
\end{table*}
\section{Methodology}
\label{sec:method}

We extend the existing \textit{SyNERGY} framework to collect power measurements and use it for developing energy predictive models. \autoref{synergy}, represents a high-level overview of the extended framework. The input to the deep learning software framework is the ConvNet model specification, the pre-trained weights from Caffe2's model repository and the input image. Inferences are executed in the deep learning software ecosystem with the necessary back-end acceleration libraries on the chosen hardware. The \textit{code annotations} supply the information of a layer's beginning and end times. The host machine remotely collects data  such as the annotations and power measurements from the target hardware. The target device runs the actual inference. The details of the target platforms in this study are provided in \autoref{platform}. Next, we describe the steps necessary to set up the software tools on both the host and target systems.

\subsection{Tools for remote monitoring on the host and Caffe2 on target mobile devices. }
The host system runs ARM Streamline version: 5.28.1, Linux 64-bit version that is compiled using the sources from the DS-5 Development studio \citep{ARMstreamline}. To facilitate this communication between the host machine and the target machine, we need the \textit{gator daemon} which communicates with the host's Streamline and the \textit{gator driver} as a loadable kernel module \citep{ARMstreamline}. The Caffe and Caffe2 binary can be built directly for the Jetson TX1 while for the Snapdragon we cross-compile the Caffe2's Android binary using the \textit{android-ndk-r16} toolchain. To integrate ARM Streamline with Caffe and Caffe2 we use the Streamline annotation library. We identify the specific functions that call each layer in the software stack in Caffe and Caffe2 and place the code annotations. This includes Caffe's \textit{net.cpp} and Caffe2's \textit{net$\_$simple.cpp}. We use the older Caffe with OpenBLAS support as done in the SyNERGY framework \citep{synergy}.
\subsection{Power measurement set up on the target}
The Jetson TX1 development comes with an on-board TI-INA3221x power sensor chip that has to be enabled during kernel source cross-compilation\footnote{List of kernel modifications can be found in \url{https://github.com/ARM-software/gator}} along with enabling its entry in the \textit{device tree binaries (dtb)}. This power sensor provides system-level power, CPU-level power and GPU-level power in mW.  We use the system-level power to the SoC as this accounts for the power due to the processing core, DRAM memory and peripherals. The power measurements are gathered with the default \textit{interactive} Linux governor. The Snapdragon 820 development board comes with on-board power pins. To capture energy measurement, we use the ARM energy probe to provide system-level power for the SoC.
%Once power measurements are obtained, the total energy for the inference $E_{inference}$ is calculated offline. 

%Similar to the TX1, the Snapdragon 820's kernel has to be cross-compiled from source to enable profiling support. 
%However, unlike the power sensor chip available on the Jetson TX1.

In our study, the power values are collected as the inference phase executes on either the target CPU (single-threaded) or the GPU. We report the execution time per image (sec/image) and energy per image (mJ/image) averaged over 5 separate runs for single image inferences. The power sampling rate is fixed to 1 kHz. To extract per layer measurements the execution profile with time-stamped code-annotations are aligned to the time-stamped power profile. We then use the extracted power measurement to calculate the energy $E_{inference}$ consumed over the time duration as per \autoref{eq1}.
\begin{equation} \label{eq1}
E_{inference}=\sum_{i=0}^T  P_{i+1} \times dt 
\end{equation}
where $P_{i+1}$ is the $(i+1)^{th}$ power sample over the time duration $dt= t_{i+1}-t_{i}$ and $T$ is the total execution time of inference application.

\begin{table*}[t]
\centering
\caption{GoogleNet per layer-type breakdown of energy and time for Eigen Library on Cortex-A57 and Kryo CPU}
\label{googlenet1}
\resizebox{0.8\textwidth}{!}{%
\begin{tabular}{|l|l|l|l|l|l|l|l|l|}
\hline
\multirow{2}{*}{\textbf{GoogleNet-Caffe2} } & \multicolumn{4}{l|}{\textbf{Eigen - Jetson TX1}} & \multicolumn{4}{l|}{\textbf{Eigen - Snapdragon 820}} \\ \cline{2-9} 
 & \textbf{Energy (mJ)} & \textbf{Time (sec)} & \textbf{\begin{tabular}[c]{@{}l@{}}Avg.\\ energy (\%)\end{tabular}} & \textbf{\begin{tabular}[c]{@{}l@{}}Avg. \\  time (\%)\end{tabular}} & \textbf{Energy (mJ)} & \textbf{Time (sec)} & \textbf{\begin{tabular}[c]{@{}l@{}}Avg.\\ energy (\%) \end{tabular}} & \textbf{\begin{tabular}[c]{@{}l@{}}Avg. \\  time (\%)\end{tabular}} \\ \hline
\textbf{Conv} & 7856.84 $\pm$ 457.2 & 1.7354 $\pm$ 0.1 & 84.44 & 84.14 & 842.66 $\pm$ 95.86 & 0.3866 $\pm$ 0.03 & 77.64 & 77.28 \\ \hline
\textbf{Fc} & 156.35 $\pm$ 22.9 & 0.0344 $\pm$ 0.0 & 1.68 & 1.66 & 18.20 $\pm$ 6.04 & 0.0082 $\pm$ 0.0 & 1.67 & 1.63 \\ \hline
\textbf{MaxPool} & 504.41 $\pm$ 46.9 & 0.1126 $\pm$ 0.01 & 5.42 & 5.45 & 59.87 $\pm$ 3.5 & 0.0284 $\pm$ 0.0 & 5.51 & 5.67 \\ \hline
\textbf{LRN} & 637.29 $\pm$ 22.4 & 0.1468 $\pm$ 0.0 & 6.84 & 7.11 & 137.62 $\pm$ 10.8 & 0.0646  $\pm$ 0.0 & 12.68 & 12.91 \\ \hline
\textbf{ReLU} & 93.37 $\pm$ 14.4 & 0.021 $\pm$ 0.0 & 1.00 & 1.01 & 23.14 $\pm$ 15.4 & 0.0106 $\pm$ 0.0 & 2.13 & 2.11 \\ \hline
\textbf{AveragePool} & 3.59 $\pm$ 2.01 & 0.0008 $\pm$ 0.0 & 0.03 & 0.03 & 0.28 $\pm$ 0.0 & 0.0002 $\pm$ 0.0 & 0.02 & 0.03 \\ \hline
\textbf{Concat} & 52.12 $\pm$ 5.4 & 0.0114 $\pm$ 0.0 & 0.56 & 0.55 & 3.00 $\pm$ 2.54 & 0.0014 $\pm$ 0.0 & 0.27 & 0.27 \\ \hline
\textbf{Dropout} & 0 & 0 & 0 & 0 & 0 & 0 & 0 & 0 \\ \hline
\textbf{Softmax} & 0 & 0 & 0 & 0 & 0.48 & 0.0002 & 0.04 & 0.03 \\ \hline
\textbf{Average Total} & 9304.00 $\pm$ 507.1 & 2.0624 $\pm$ 0.0 & 100 & 100 & 1085.28 & 0.5002 & 100 & 100 \\ \hline
\end{tabular}
}
\end{table*}

\begin{table}[t]
\centering
\caption{GoogleNet per layer-type breakdown of energy and time for OpenBLAS - Cortex-A57}
\label{googlenet2}
\resizebox{0.45\textwidth}{!}{%
\begin{tabular}{|l|l|l|l|l|}
\hline
\multirow{2}{*}{\textbf{GoogleNet-Caffe}} & \multicolumn{4}{l|}{\textbf{OpenBLAS Jetson TX1}} \\ \cline{2-5} 
 & \textbf{Energy (mJ)} & \textbf{Time (sec)} & \textbf{\begin{tabular}[c]{@{}l@{}}Avg. \\ energy (\%)\end{tabular}} & \textbf{\begin{tabular}[c]{@{}l@{}}Avg. \\  time (\%)\end{tabular}} \\ \hline
\textbf{Conv} & 4883.26 $\pm$ 818.6  & 1.52 $\pm$ 0.2 & 62.36 & 62.66 \\ \hline
\textbf{InnerProduct} & 138.96  $\pm$ 30.7 & 0.04  $\pm$ 0.01 & 1.77 & 1.76 \\ \hline

\textbf{Pooling} & 761.83 $\pm$ 92.2 & 0.23 0.03 & 9.72 & 9.82 \\ \hline
\textbf{LRN} & 1716.78 $\pm$ 43.1 & 0.52 $\pm$ 0.01 & 21.92 & 21.53 \\ \hline
\textbf{ReLU} & 178.86 $\pm$ 35.9 & 0.05 $\pm$ 0.01 & 2.28 & 2.29 \\ \hline
\textbf{Split} & 9.00 $\pm$ 6.0 & 0.002 0.0  & 0.11 & 0.11 \\ \hline
\textbf{Concat} & 128.59 $\pm$ 22.5 & 0.03  $\pm$ 0.0 & 1.64 & 1.63 \\ \hline
\textbf{Dropout} & 0.6798  $\pm$ 1.5 & 0.0002  $\pm$ 0.0 & 0.008 & 0.008 \\ \hline

\textbf{Softmax} & 1.85  $\pm$ 1.6 & 0.0006  $\pm$ 0.0 & 0.02 & 0.02 \\ \hline
\textbf{Average Total} & 7830.07 & 2.43 & 100 & 100 \\ \hline
\end{tabular}}
\end{table}

\section{Layer-type Energy and Performance Measurements}
\label{sec:energymeasurements}
In this section, we use SyNERGY to provide empirical time and energy measurements for the overall inference as well as finer-grained layer-type for an example ConvNet model (in our case, GoogleNet). % is when Layer-type we sum the energy and execution time of each layer  into a layer-type category. 
Layer-type for the convolutional layers is when
we group \textit{individual} convolutional layers regardless of its specifications (for example, $11 \times 11$, $3 \times 3$ or $1 \times 1$) into a broader category of Conv.  

%Our measurements are reported for three cases of software-hardware combinations: Eigen-TX1, Eigen-Snapdragon820 and OpenBLAS-TX1 for GoogleNet model. 
We summarize our findings for Eigen-TX1 and Eigen-Snapdragon820 in \autoref{googlenet1}, and  OpenBLAS-TX1 in \autoref{googlenet2}. In the case of Caffe2, the pooling layer is further split into two types: MaxPool and AveragePool while in the case of the original Caffe framework both versions are grouped under the Pooling category. Certain layers like dropout and softmax execute too quickly and are too small to be captured. We also report the average percentage of energy and time of each layer when compared to the total energy and time. If we compare the total inference energy in all three software and hardware cases, the combination with the least amount of energy per inference is the Caffe2's Eigen-Snapdragon820. However, comparing the energy of the Conv layer for the Jetson TX1 with both software backends, we observe that OpenBLAS consumes $1.6\times$ less energy than Eigen.

If we compare the energy of the different layers, we observe that the Conv layer contributes the most to the total energy consumed in the inference phase ($62 - 85 \%$) across all three combinations. 
%This is due to the cost of executing each layer on each system as well as  the number of convolutional layers present in a model. %There are considerable efforts to optimize these issues \citep{hal, fused, squeezenet} and we focus on obtaining the energy-use in the context of such optimizations.
The pooling ($5-9 \%$) and LRN ($6 -21 \%$) rank second making them good candidates for optimization. 
In this section, we use the energy measurements obtained from SyNERGY to evaluate the energy consumption of a ConvNet model to identify energy bottlenecks and perform comparative analysis for different software and hardware systems. In  Section~\ref{overall}, we show how predictive models based on energy can be used to perform such comparative analysis. Finally, this \textit{layer-type} abstraction will be useful in the next sections, where we focus on building predictive models at the granularity of layer-types and compare it with previous approaches that use individual layers for predictive models.

\section{Feature Selection}
\label{sec:feature-selection}
 Previous work such as Neuralpower \citep{neuralpower}, build layer-wise predictive models by using complex features extracted from individual layers. For example, higher order terms for kernel size, input volume and others. While other works \cite{synergy} use simple aggregate algorithmic features (we refer to this as layer-type), for example, an aggregate MAC count to build predictive models for the convolutional layers. 
 Therefore, in this section, we first aim to evaluate algorithmic features (highlighted in bold in Section 2) extracted for individual layers to build predictive model in terms of predictive model accuracy and complexity. Our feature selection is based on standard techniques of \textit{best subset selection}  using metrics such as  \textit{Bayesian Information Criterion} (BIC) \citep{james2013introduction}. These methods are typically used to evaluate the  trade-off in model complexity and accuracy as features are added to the model. We refer to model complexity as the number of features in the final predictive model. 

\begin{figure}[h]
\centering
\subfloat[Linear features]{\includegraphics[width=0.5\columnwidth]{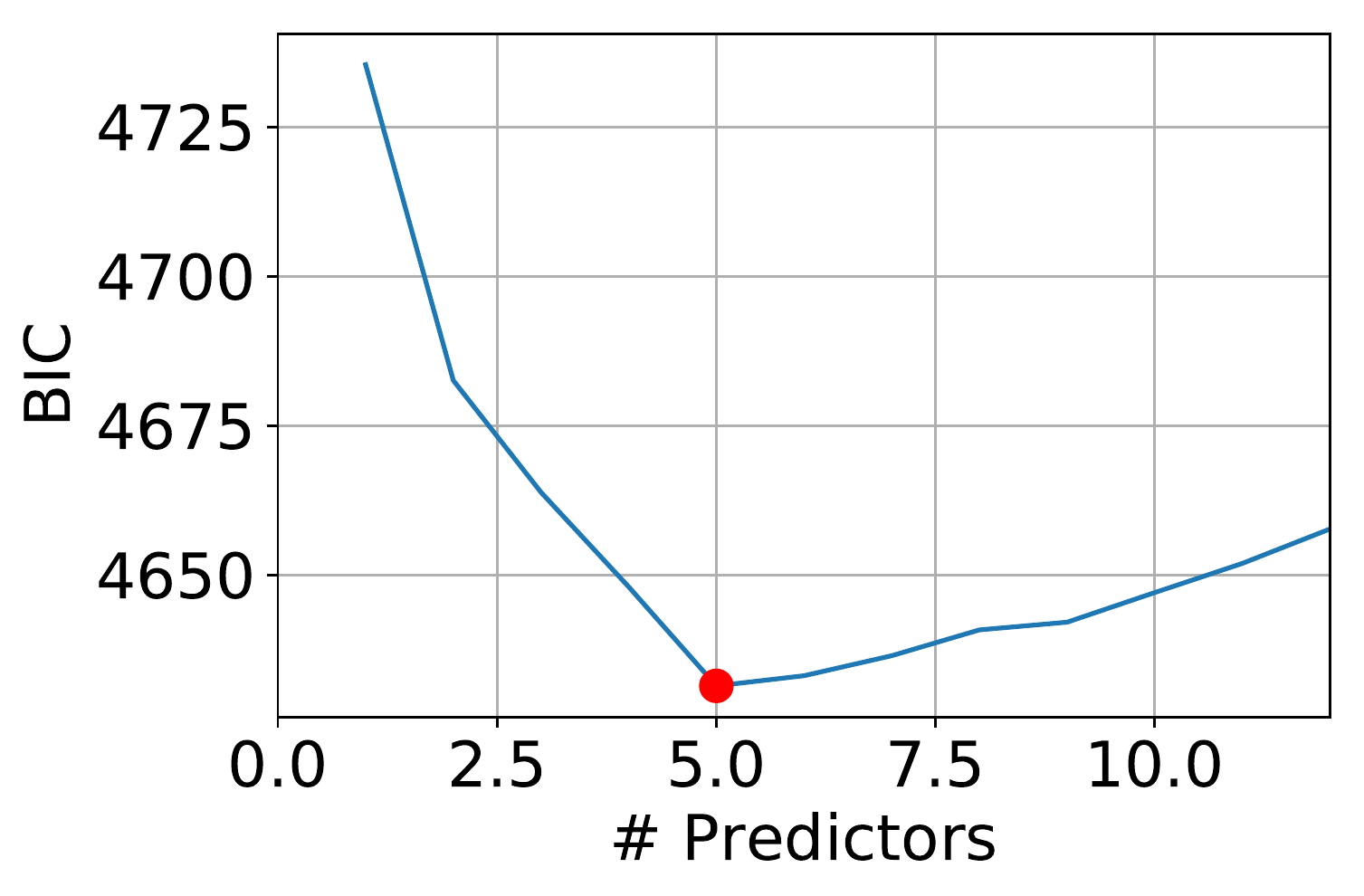}
\label{fig:first-bic}}
~
\subfloat[Non-linear features]{\includegraphics[width=0.5\columnwidth]{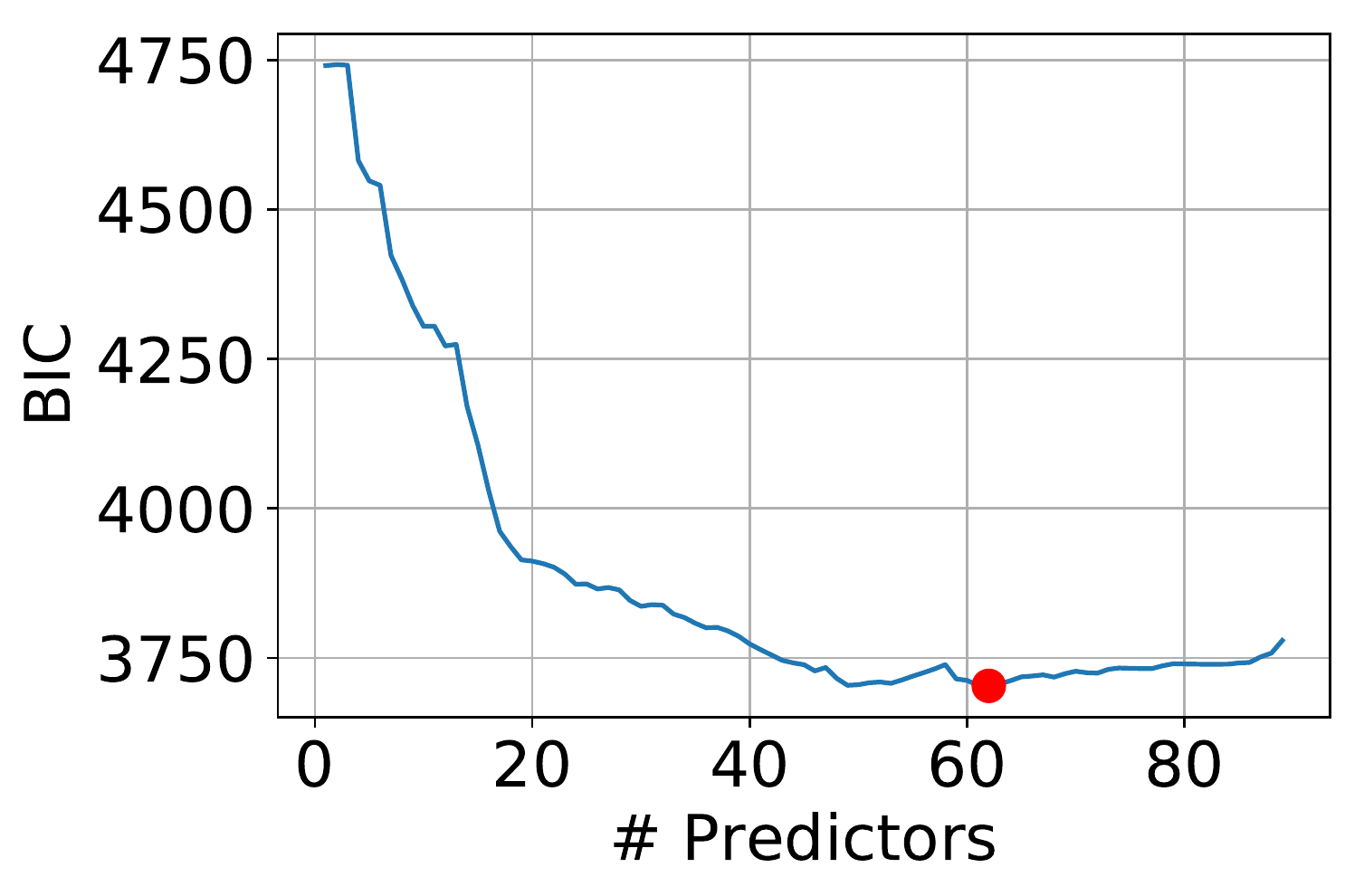}
\label{fig:second-bic}}
\caption{Subset feature selection. Lower BIC is better.}
\label{bic-fig}
\end{figure}

We demonstrate this feature analysis for all the convolutional layers of all 12 target ConvNets executing on the CPU of Jetson TX1 with OpenBLAS backend.
The linear features (or degree d=1) for these layers include: kernel shape, padding, stride, $I_x$ (same as $I_y$), $O_x$ (same as $O_y$), $Oz$, $I_z$, input size, output size, weights, data volume and MAC.  In this case, each convolutional layer has a set of 12 features. The target response is the energy for an individual layer. Figure~\autoref{fig:first-bic}, shows that a model with 5 features (indicated by red circle as the lowest BIC) would be a good a choice. To model non-linear features, we extended the linear feature set to consist of higher order polynomial terms and cross terms (of degree $d=2$)  for these features (this includes, $kernel^2$, $kernel\times stride$ and others).  For a degree 2 model the model with lowest BIC has a model complexity of  62 features, as shown in Figure~\autoref{fig:second-bic}.

\begin{table}[t]
\caption{BIC subset selection for the Convolutional layer.}
\label{bic}
\centering
\resizebox{1.0\columnwidth}{!}{
\begin{tabular}{|l|l|l|l|l|l|}
\hline
\textbf{\begin{tabular}[c]{@{}l@{}}Polynomial \\ degree (d)\end{tabular}} & \textbf{\begin{tabular}[c]{@{}l@{}}Model \\ Complexity \\ (m)\end{tabular}} & \textbf{Model} & \textbf{\begin{tabular}[c]{@{}l@{}}BIC\\ (lower is\\ better)\end{tabular}} & \textbf{Features} & \textbf{\begin{tabular}[c]{@{}l@{}}Relative \\ comparison \\ to MAC model\end{tabular}} \\ \hline
1 & 1 & \textit{MAC model} & 4753 & MAC & 1 \\ \hline
1 & 5 & \textit{Best linear model} & 4631 & \begin{tabular}[c]{@{}l@{}}Kernel, Stride, Ox, \\ Output Volume, MACs\end{tabular} & 2.63 \\ \hline
2 & 62 & \textit{\begin{tabular}[c]{@{}l@{}}Best non-linear \\ model\end{tabular}} & 3703 & \begin{tabular}[c]{@{}l@{}}62 linear \&  non-linear \\ features\end{tabular} & 28.35 \\ \hline
\end{tabular}}
\end{table}

In order to understand predictive model should yield greater predictive accuracy in \autoref{bic}, we make a relative comparison of a single feature-based model using MAC to the predictive model using the best combination of linear features and to the predictive model with best combination of linear and non-linear features obtained from the previous step. Based on the relative comparison, a single feature MAC model is found to be within 3\% of the best linear feature model and within 29\% of the best non-linear feature model.  Therefore, to get a highly accurate predictive model which is indicated by a lower BIC, the number of features extracted for individual layers would be in the order of 62 non-linear features.  In the next section, we compare regression-based models based on individual features with predictive models based on layer-type features on the basis of predictive accuracy to determine whether higher complexity models indeed offer better accuracy when compared to lower-complexity models.

\begin{table*}[t]
\centering
\caption{Comparison of energy predictive accuracy and model complexity}
\label{compareneural}
\resizebox{0.7\textwidth}{!}{%
\begin{tabular}{|l|l|l|l|l|l|}
\hline
\multirow{2}{*}{\textbf{Dataset}} & \multirow{2}{*}{\textbf{OpenBLAS-TX1}} & \multirow{2}{*}{\textbf{Model type}} & \multirow{2}{*}{\textbf{\begin{tabular}[c]{@{}l@{}}10-Fold cross \\ validation accuracy\end{tabular}}} & \multirow{2}{*}{\textbf{\begin{tabular}[c]{@{}l@{}}Polynomial \\ degree (d)\end{tabular}}} & \multirow{2}{*}{\textbf{\begin{tabular}[c]{@{}l@{}}Model \\ complexity (m)\end{tabular}}} \\
 &  &  &  &  &  \\ \hline
\begin{tabular}[c]{@{}l@{}}Overall Conv\\ model\end{tabular} & {$MAC_{sum}$ model} & Energy & 81.84 $\pm$ 7.8 & 1 & 1 \\ \hline
\multirow{5}{*}{\begin{tabular}[c]{@{}l@{}}Individual \\convolutional \\ layer model\end{tabular}} & \textit{MAC model} & Energy & 67.02 $\pm$ 11.91 & 1 & 1 \\ \cline{2-6} 
 & \textit{\begin{tabular}[c]{@{}l@{}}Best linear model\end{tabular}} & Energy & 72.83 $\pm$ 10.7 & 1 & 5 \\ \cline{2-6} 
 & \textit{\begin{tabular}[c]{@{}l@{}}Best non-linear model\end{tabular}} & Energy & 79.58 $\pm$ 13.03 & 2 & 32 \\ \cline{2-6} 
 & \multirow{2}{*}{\textit{NeuralPower} \citep{neuralpower}} & Runtime & \multirow{2}{*}{77.48 $\pm$ 21.21} & 2 & 4 \\ \cline{3-3} \cline{5-6} 
 &  & Power &  & 2 & 17 \\ \hline
\end{tabular}}
\end{table*}

\subsection{Analysis of Model Accuracy \& Complexity }
\label{sec:analysis}
Based on our analysis of features in the previous section, we build regression based models trained using the standard \textit{supervised learning} approach in machine learning \citep{james2013introduction}.  Cross validation is performed 10 times and the convolutional layers used in train and test sets are in the ratio 80:20.
The regression-based model layer-wise predictive models is given by \autoref{econveq}. Similar predictive models can be built for other layers to give the overall energy of the inference as the sum of the predictions from all the layer-wise predictive models, as given by \autoref{einfeq}.
\begin{equation}
\label{econveq}
        \hat{E}^{layer}=x_{1} \times feature_{1}^{1} + ... + x_{d_m} \times feature_{m}^{d_m}
\end{equation}

\begin{equation}
\label{einfeq}
     \hat{E}^{inference} = \hat{E}^{conv} + \hat{E}^{pool} ... +\hat{E}^{layer}
\end{equation}
where $d_m$ represents the degree of the $m^{th}$ algorithmic feature.

As described in the previous sections, we have two types of predictive models based on the type of features: layer-wise predictive models use features aggregated across layers while individual layer models use features from every layer.
The individual convolutional layer models are of four categories, as given in \autoref{compareneural}: a single feature model (MAC) without summation counts (as done in \citep{synergy}), a model with the best BIC for linear features (best linear),  a model with the best BIC for non-linear features (best non-linear) and finally, we compare with a previous work, Neuralpower \citep{neuralpower}. NeuralPower is  based on predictions from run-time and power estimation models, to get an estimate of time and power, and subsequently energy, for individual layers. For this, we use the code provided by NeuralPower\footnote{\url{https://github.com/cmu-enyac/NeuralPower}}.

By comparing the models based on individual layer features, as summarised in \autoref{compareneural}, we find that using a larger set of complex features does not provide a massive boost in accuracy compared with the use of  simpler features, as indicated by the results from \autoref{bic}; for example, compare the data for complex models such as the Best non-linear model and NeuralPower in \autoref{compareneural} with that for the simpler models such as the MAC model and Best linear model.
Furthermore, for a single hardware and software configuration and a single split of their dataset, Neuralpower reports an overall accuracy of 97.21\% (based on the Root-mean-squared-percentage-error (RMSPE)). When using Neuralpower on our dataset,
we observe similar high accuracies for  certain splits (for example,
considering the upper bound we get 77.47+21.21=98.69\%). However, this behaviour is not consistently observed across other splits of train and test sets as done in our experimental evaluation. Our results indicate a mean and variance of the accuracy of 77.47$\pm$21.21\% in \autoref{compareneural},  across different splits of training and test sets.

On further analysis of the results of NeuralPower, we observe that certain ConvNet models are over-predicted while others are under-predicted when using two different predictive models - the runtime and power models - leading to a {\em cancellation effect} when calculating energy to give an overall high accuracy which may be misleading.  However, we do not observe such cancellation effects when using a simple $MAC_{sum}$ model trained directly on energy use information. In addition, the $MAC_{sum}$ model offers a higher accuracy and lower variance compared to using models based on individual layers (See Column 4 of \autoref{compareneural}).
%In the next section, we compare the predictions obtained using the different predictive models with the measured energy data. We find that individual layer predictive models suffer from certain layers being over-predicted while others are under-predicted; this can lead to a {\em cancellation effect} when summing the predictions leading to a high overall accuracy, which may be misleading. However, we show that the simple $MAC_{sum}$ model does not suffer this cancellation affect in its summation process which takes place {\em prior} to prediction. Instead, the $MAC_{sum}$ model yields a higher and more stable predictive accuracy but with model complexity 4$\times$ and 17$\times$ lower compared to the  predictions of runtime and power, based on summing individual layers, obtained using NeuralPower. 

Given this behaviour, we conclude that predictive models based on $MAC_{sum}$ offers a good first approximation to estimate the energy consumed in a ConvNet in terms of model complexity and accuracy. The $MAC_{sum}$ model yields a higher and more stable predictive accuracy but with model complexity 4$\times$ and 17$\times$ lower compared to previous approaches that use data from individual layers (Column 6 of \autoref{compareneural}). Moreover, MAC (or more generally an operation count, Op) is a universal feature that can be extracted for other layer-types (see Section~\ref{sec:overall-energy-prediction}). Therefore in  next section, we extend the $MAC_{sum}$ approach to other layer-types to get an overall estimate of the energy consumed by the deep learning model.

\section{Overall Inference Energy}
\label{sec:overall-energy-prediction}
In this section, we extend our method to construct per layer energy prediction models for the Conv, Fc and pooling layers using MAC count (or, equivalently, OpCount for pooling layers). 
To make a prediction for an entire ConvNet model, the training and test sets are split based on the ConvNet models themselves during cross-validation. This ensures that for given test ConvNet all its layers are present only in the test set. 
%Initially, we compare the use of linear regression-based model using MAC or (OpCount for pooling layer) with higher order polynomial-based models of degree = 2  using the same feature. For the sake of completeness, we show the regression coefficients of both linear and polynomial models for the Conv, Fc and pool layer in Figure 3 (left) and Figure 3 (right) and Figure 4 espectively.
%To compare the performance of both the regression models we ensure they are evaluated on the same set of test examples in each of the 10-fold split.
The predictive models are evaluated in terms of their \textit{relative accuracy}, given by \autoref{relerr2}, which quantifies the relative performance of the predictor with respect to the baseline measured energy value \citep{synergy}. We average the relative test errors across all test examples and across all folds of data.

\begin{equation}
\label{relerr2}
  Rel.Acc (\%) = 100-\Big(\frac{|predicted - measured|}{measured} \times 100\Big)
\end{equation}

\subsection{Layer-type predictions}
\label{convlayers}
\begin{figure*}[h]
\centering
\subfloat[Eigen-Snapdragon820]{\includegraphics[width=0.6\columnwidth]{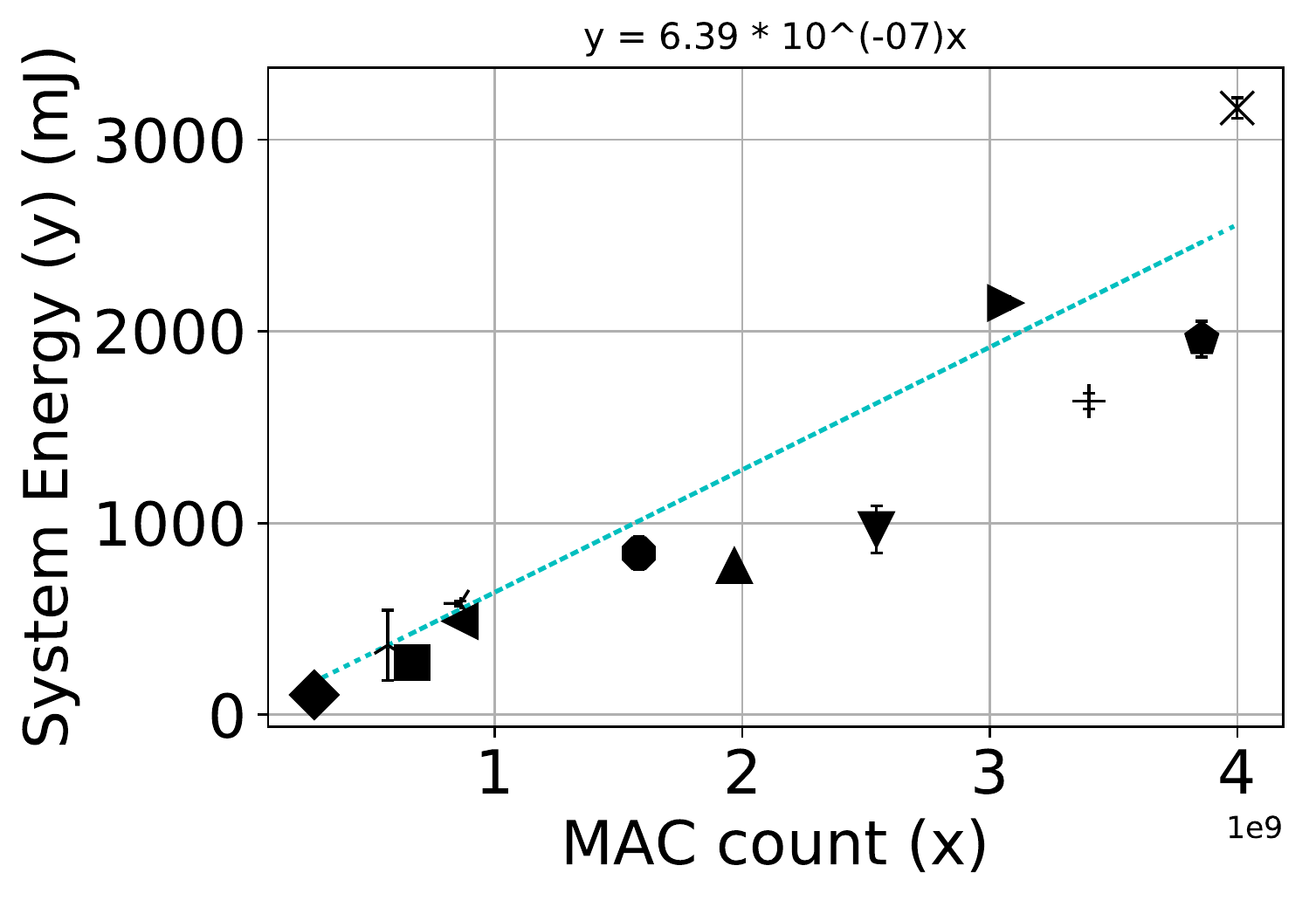}
\label{fig:first}}
~
\subfloat[Eigen-TX1]{\includegraphics[width=0.6\columnwidth]{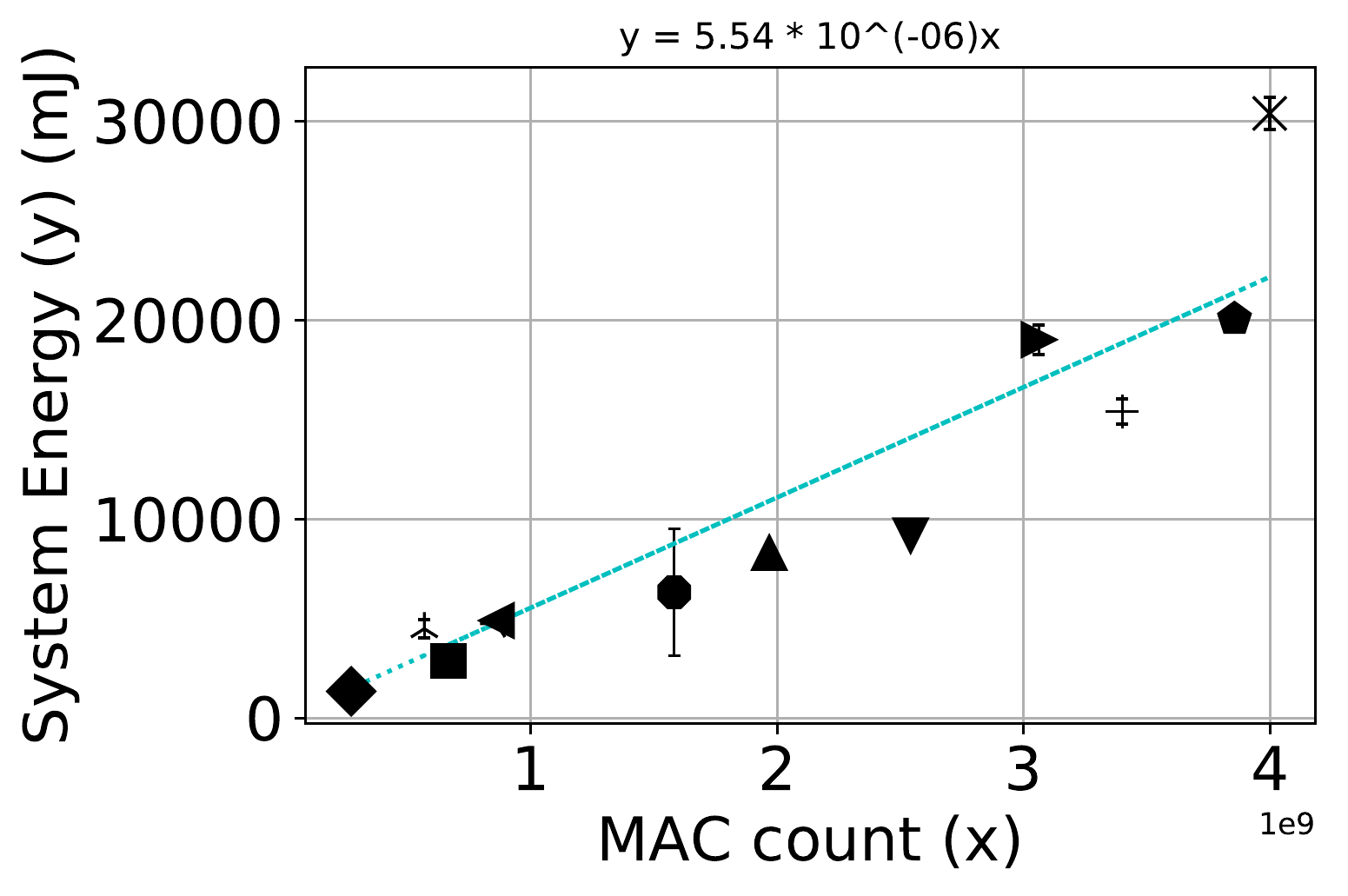}
\label{fig:second}}
~
\subfloat[OpenBLAS-TX1]{\includegraphics[width=0.8\columnwidth]{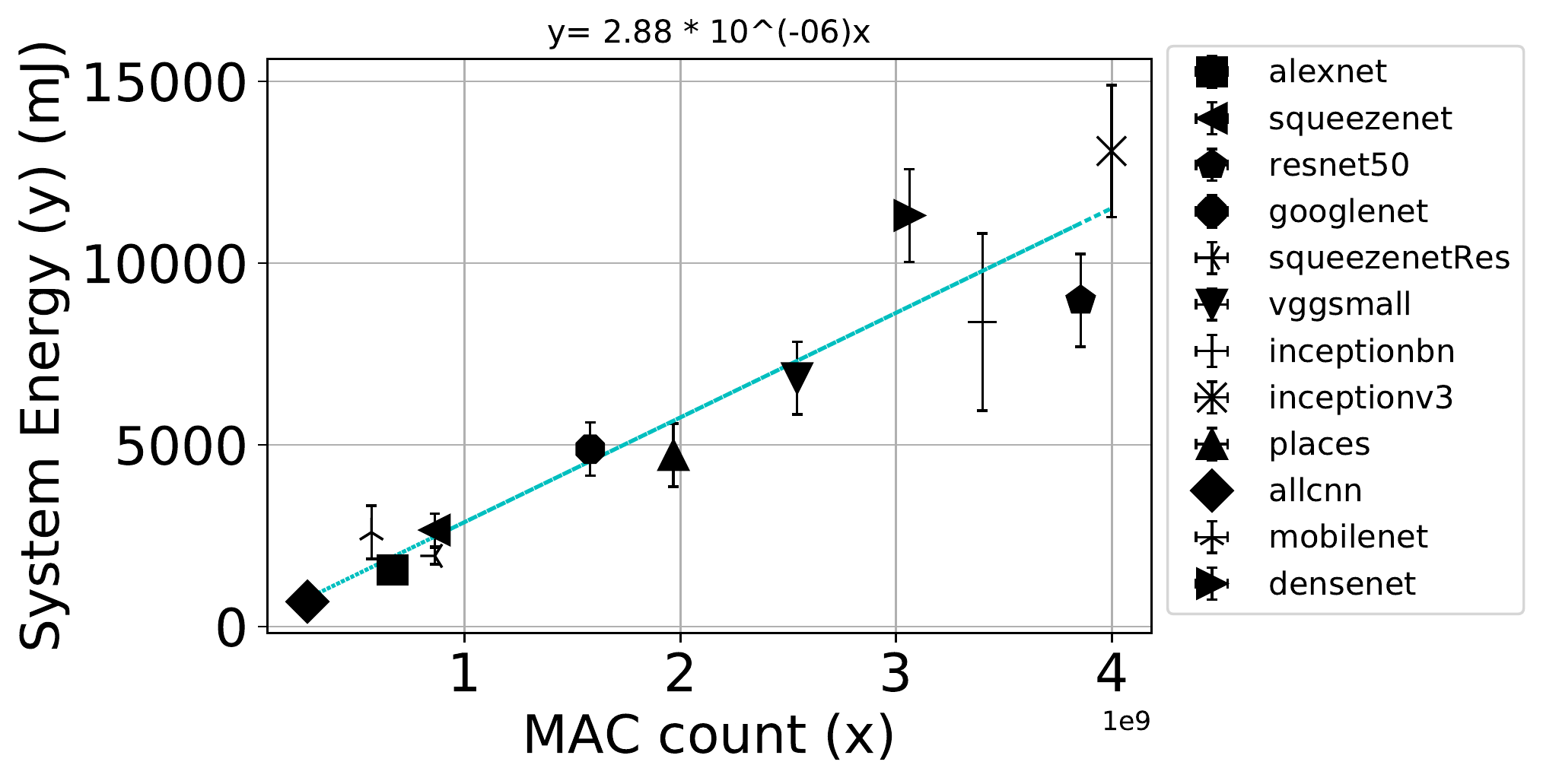}
\label{fig:third}}
\caption{Linear regression-based energy prediction models using MAC for the Conv layer.}
\label{fig:conv}
\end{figure*}

From \autoref{fig:conv}, we plot the linear regression model over the data points in our dataset. We observe that the relative positions of each data point in all three cases (that is, Eigen-Snapdragon820, Eigen-TX1 and OpenBLAS-TX1) follow a similar trend. At the bottom left corner, we observe models with low MAC count and low energy use, for example, squeezenet and mobilenet. It is interesting to observe that smaller sized models, in terms of number of parameters, do not always result in better energy use. For example, resnet50 outperforms inceptionv3 in terms of energy use in all three cases despite being roughly the same size (see \autoref{model}). We also observe that alexnet has lower energy use than squeenzenet and mobilenet despite being approximately 3 and 2 times greater in model size \footnote{This is considering only the convolutional layers.} respectively. This is because the latter models use smaller kernel shapes such $1 \times 1$ and $3 \times 3$ to reduce the number of parameters in the model. However, these require optimized software routines for small kernel shapes to exploit the resources on a system effectively. Our results for the relative accuracy of the Conv predictive models on the test set is tabulated in \autoref{regression}. In all three cases, we find that the linear regression model using solely MAC count as an input feature achieves a test accuracy between 75\% to 82\%.

\begin{table}[t]
\centering
\caption{Per layer-type energy prediction results for all software-hardware combinations}
\label{regression}
\resizebox{0.8\columnwidth}{!}{%
\begin{tabular}{|l|l|l|}
\hline
\textbf{Layer} & \textbf{} & \textbf{\begin{tabular}[c]{@{}l@{}}10-fold cross \\ validation accuracy\end{tabular}} \\ \hline
 & \begin{tabular}[c]{@{}l@{}}Software-Hardware \\  Combination\end{tabular} & \begin{tabular}[c]{@{}l@{}}Linear Regression\\ accuracy (\%)\end{tabular} \\ \hline
\multirow{3}{*}{\textbf{Conv}} & \textit{Eigen-Snapdragon820} & 75.24 $\pm$ 12.0 \\ \cline{2-3} 
 & \textit{Eigen-TX1} & 78.16  $\pm$  6.3 \\ \cline{2-3} 
 & \textit{OpenBLAS-TX1} & 82.41  $\pm$  7.4\\ \hline
\multirow{3}{*}{\textbf{FC}} & \textit{Eigen-Snapdragon820} &  76.75  $\pm$  10.3\\ \cline{2-3} 
 & \textit{Eigen-TX1} & 64.56  $\pm$  9.3 \\ \cline{2-3} 
 & \textit{OpenBLAS-TX1} & 56.72  $\pm$  5.3 \\ \hline
\multirow{3}{*}{\textbf{Pool}} & Eigen-Snapdragon820 & 90.01  $\pm$  4.4\\ \cline{2-3} 
 & Eigen-TX1 & 82.39  $\pm$  8.1 \\ \cline{2-3} 
 & OpenBLAS-TX1 & 86.05  $\pm$  7.5 \\ \hline
\end{tabular}}
\end{table}
We aim to model the Fc layers and pooling layers using an equivalent feature to $MAC_{sum}$ count as done previously for the Conv layers. However, as observed in \autoref{model} from Section 2, there are fewer ConvNets with Fc layers, and fewer Fc layers per ConvNet model. Despite the fact that we are using a larger number of ConvNets than previous studies, the data for the Fc layers is limited. 
As seen in \autoref{regression}, the lower accuracy between 56\% to 76\% using a linear fit could be a result of insufficient data points. We could address this issue by trying to generate more points for the Fc layers by using individual Fc layers as adopted by previous approaches or generate more data points by using varying batch sizes.

For the pooling layers, we focussed on MaxPool operations as they account for more number of layers than average pooling in real ConvNets. We use the OpCount given in \autoref{compeq4}. The results in Figure 4, show that using solely the OpCount as an input feature we can obtain a linear fit with test accuracy between 82\% to 90\%.

\subsection{Overall energy predictions}
\label{overall}
In this section, we obtain an estimate for the energy for the whole ConvNet by summing the predictions from the Conv, Fc and MaxPool layer-type predictive models. We select GoogleNet, AlexNet and VGG\_CNN\_S as test data points (see \autoref{RMSPE})  because AlexNet and a variant of VGG with 16 layer were evaluated in NeuralPower \citep{neuralpower}, and GoogleNet is our running example. We use the remaining model points as the training set to form the linear model. 

\autoref{RMSPE} shows the prediction results for each layer-type Conv, Pool and Fc given by Columns 3, 4 and 5 and the overall predicted results (Column 6: Total predicted) for the inference. The measured energy for the Conv, Pool and Fc layer-type is given in Columns 7, 8 and 9 and the overall measured energy in Column 10 of \autoref{RMSPE}. Similar to the results obtained using empirical measurements in Section~\ref{sec:energymeasurements}, we find that using the predicted energy for the convolutional layers (given in Column 3) of GoogleNet, the OpenBLAS library is $1.2\times$ less energy consuming than the Eigen library for the TX1 platform. %We also compare the total predicted energy in Column 6 foe GoogleNet to show that hardware and software combination with the least amount of energy per inference is the Eigen-Snapdragon820 platform.  

Finally, we also report accuracy using RMSPE (as per the metric used in NeuralPower) and relative test accuracy (see \autoref{relerr2}). Both metrics provide similar results. We find that across the four software-hardware combinations, including mobile GPUs (CuDNN-TX1 in \autoref{RMSPE}), we achieve a significant relative test accuracy of between 76\% to 85\% using solely summation of MAC (or operation) counts as the input feature to a linear model.

\begin{table*}[t]
\caption{Aggregate Energy prediction results for Conv, pooling and Fc layers}
\label{RMSPE}
\resizebox{1.0\textwidth}{!}{%
\begin{tabular}{|l|l|l|l|l|l|l|l|l|l|l|l|}
\hline
\textbf{Software-Hardware} & \textbf{Test ConvNet} & $\hat{Conv}$ & $\hat{Pool}$ & $\hat{Fc}$ & \textbf{\begin{tabular}[c]{@{}l@{}}Total \\ predicted (mJ)\end{tabular}} & \textit{Conv} & \textit{Pool} & \textit{Fc} & \textbf{\begin{tabular}[c]{@{}l@{}}Total layer \\ measured (mJ) \end{tabular}} & \textbf{\begin{tabular}[c]{@{}l@{}}Accuracy (\%)\\``100-RMSPE"\end{tabular}} & \textbf{\begin{tabular}[c]{@{}l@{}}Rel.Test\\ Accuracy (\%)\end{tabular}} \\ \hline
\multirow{3}{*}{\textbf{Eigen-Snapdragon820}} & \textit{GoogleNet} & 604.39 & 54.96 & 9.98 & 669.33 & 842.66 & 59.87 & 18.2056 & 920.7356 & \multirow{3}{*}{81.41} & \multirow{3}{*}{83.11 $\pm$ 9.4} \\ \cline{2-10}
 & \textit{AlexNet} & 281.55 & 5.16 & 599.64 & 886.35 & 271.842 & 5.46 & 495.81 & 773.112 &  &  \\ \cline{2-10}
 & \textit{VGG\_CNN\_S} & 1306.34 & 7.89 & 815.36 & 2129.59 & 966.55 & 6.77 & 985.69 & 1959.01 &  &  \\ \hline
\multirow{3}{*}{\textbf{Eigen-TX1}} & \textit{GoogleNet} & 5783.59 & 1079.29 & 182.73 & 7045.61 & 6325.22 & 1583.87 & 156.35 & 8065.44 & \multirow{3}{*}{84.70} & \multirow{3}{*}{84.81 $\pm$ 2.2} \\ \cline{2-10}
 & \textit{AlexNet} & 2476.42 & 93.24 & 9934.27 & 12503.93 & 2875.25 & 104.94 & 11885.76 & 14865.95 &  &  \\ \cline{2-10}
 & \textit{VGG\_CNN\_S} & 10285.35 & 170.43 & 19537.94 & 29993.72 & 9177.43 & 123.86 & 16331.49 & 25632.78 &  &  \\ \hline
\multirow{3}{*}{\textbf{OpenBLAS-TX1}} & \textit{GoogleNet} & 4534.3 & 1272.75 & 276.89 & 6083.94 & 4883.26 & 765.16 & 171.4 & 5819.82 & \multirow{3}{*}{71.61} & \multirow{3}{*}{76.24 $\pm$ 9.02 } \\ \cline{2-10}
 & \textit{AlexNet} & 1908.68 & 67.34 & 17318.97 & 19294.99 & 1562.12 & 86.47 & 11884.87 & 13533.46 &  &  \\ \cline{2-10}
 & \textit{VGG\_CNN\_S} & 7285.57 & 121.1 & 19536.91 & 26943.58 & 6837.69 & 213.83 & 28472.67 & 35524.19 &  &  \\ \hline
\multirow{3}{*}{\textbf{CuDNN-TX1}} & \textit{GoogleNet} &  1471.22 & 484.53  & 52.04& 2007.79 &  2579.81&409.51   &  84.16& 3073.48
 & \multirow{3}{*}{77.54} & \multirow{3}{*}{82.43 $\pm$ 17.0} \\ \cline{2-10}
 & \textit{AlexNet} & 619.30  & 35.56 & 2959.56 & 3614.42  & 527.27 &38.26 & 3033.99 & 3599.525
&  &  \\ \cline{2-10}
 & \textit{VGG\_CNN\_S} & 2363.91 & 64.70& 4988.58 & 7417.19
 & 1362.50& 78.48 & 4864.99 & 6305.97
 &  &  \\ \hline
\end{tabular}}
\end{table*}

\section{Related Work}
\label{sec:relatedwork}

To enable efficiency in deep learning algorithms, software and hardware will require better understanding in the energy use of deep learning models. %In particular, modelling the energy use on existing hardware can provide researchers a way to explore energy and performance trade-offs before actual execution.
This section covers related work in the areas of performance and energy benchmarking, and performance and energy modelling. 

\textbf{Performance and energy benchmarking:}
Performance or execution time is used as metric to evaluate deep learning models on existing desktop and server systems as done in Fathom \citep{fathom}. These studies are representative of execution environments with powerful processors and availability of larger memories which is not typically representative of resource constrained low-powered devices. Our work instead provides \textit{both} performance and energy use of 12 representative ConvNet models when executing on resource constrained mobile systems and identifies energy bottlenecks at a fine-grained level. %By using energy as a metric alongside performance we open up the opportunities to study power management techniques that could be applied at run-time to optimize the application's energy use \citep{synergy} targeted at deep learning applications. 

Recent energy benchmarking efforts, such as BenchIP \citep{benchip} have emerged to understand the energy use of deep learning applications across different types of hardware systems. The authors develop a benchmark suite of single layers and full ConvNet models and is aimed at evaluating different hardware systems. However, it is unclear how usable this framework would be for measurement and modelling studies described in this paper as it is yet to be open-sourced. 

\textbf{Performance and energy modelling:}
To overcome the requirement of having to execute every model to measure its performance, recent studies \citep{modeling} have focussed on modelling the execution time and resource usage for only the convolutional layers in a ConvNet model. They use matrix multiplication as a major component in a convolutional layer and execute different matrix sizes in isolation to model its performance and resource use. The authors identify that such isolation failed to capture the dependencies  between layers during actual inference runs leading to an over-estimate in the prediction of execution time compared to actual execution time. Our work instead captures the energy use of the layers in the context of the execution environment of an entire inference and uses it to build predictive models. 

Early studies \citep{surveyefficientdnns}, relied on counting the number of weights of the deep learning model and energy look-up tables for estimating the energy cost of DRAM memory accesses during the inference phase on specialized hardware. 
%For spatial architectures such as ASICs, energy estimation models that the model computation and memory exists\citep{yang2016designing,eye}. These models are built by counting the number of accesses at each level in the memory hierarchy of the spatial architecture and attributing each access to an energy cost using an energy conversion table. 
However, such estimation models for deep learning on general purpose processing processors such as CPUs and GPUs have only recently emerged (for example, a mobile CPU \citep{synergy} and desktop GPUs \citep{neuralpower}.)
Our work instead builds upon the former that models the energy consumption using platform-specific performance counter information. Specifically, we build predictive models at the application-level using \textit{platform-agnostic} neural network features. 

Our work also shares similarity to NeuralPower \citep{neuralpower} that builds predictive models for a desktop GPU. However, we differ in three main aspects. First, NeuralPower develop per layer power and runtime prediction models on desktop GPUs such as Titan X GPU to predict the energy for 5 ConvNet test models. Our work focusses on empirical power measurements obtained in a resource constrained mobile devices and comprehensively evaluates 12 representative ConvNets. Second, NeuralPower does not provide an analysis on how to select features used to build their predictive models. We use statistical analysis to select dominant input features extracted from the algorithm. Third, although the average energy prediction accuracy reported by NeuralPower appears high, it can not be replicated on a different set of ConvNets as shown in our results in Section 6.
\section{Discussion}
\label{sec:discussion}
The overhead of the power measuring software, introduced by the \textit{gator daemon}, executing on the target device is negligible, approximately 3\% \citep{ARMstreamline}. Our data collection phase on each platform takes around 5 minutes for all 12 ConvNets. Our feature selection process takes less than a minute. Predictive model training and testing takes approximately 5ms. This low overhead is beneficial, as for a new software-hardware configuration, we only pay this cost once and use a few ConvNets to approximate the energy use on the platform. 

Finally, the predictive models in our work are built at the layer-level and any optimizations to accelerate a layer, such as fused-layer implementations \citep{fused}, is typically done below this level of abstraction, and is thus automatically covered. 
\section{Conclusions and Future Scope}
\label{sec:conclusion}

Deep neural network inference is becoming increasingly popular on low-power mobile devices. 
In this work, we focus on building energy predictive models by thoroughly investigating the impact of choosing application-level features on the final predictive model accuracy and complexity. 

To support building of predictive models we extended the SyNERGY- a framework for gathering energy measurements on different mobile devices. 
%We comprehensively evaluate 12 representative convolutional neural networks across different software and hardware combinations (specifically, mobile CPUs and GPU): Eigen-Snapdragon820, Eigen-TX1, OpenBLAS-TX1 and CuDNN-TX1. 
We compare two types of predictive models found in the literature - based on features selected for layers at different levels - individual layers and layer-type.
Our analysis using subset feature selection techniques for individual layer models indicate that highly complex features are required to achieved greater predictive accuracy. However unlike the results of previous works, we find that predictive models based on layer-type features (for example, summation of  operation counts) offer a  better model complexity of 4 to 32 times less than  models using individual layer features for a similar average accuracy ($\approx 82\%$). We further demonstrate that such an approach can be extended to other layer-types with an accuracy of between 76\% to 84\% using solely summation counts of MAC or Op counts as the input feature to a linear model across different mobile hardware and software combinations (specifically, mobile CPUs and GPU): Eigen-Snapdragon820, Eigen-TX1, OpenBLAS-TX1 and CuDNN-TX1. 

As future work, we aim to extend our modelling studies to layers found in other types of deep neural networks such as Recurrent Neural Networks, other devices and further explore non-linear modelling strategies.
\begin{acks}
This  research  was  conducted  with  support  for  C. Rodrigues and G.D. Riley from the IS-ENES2 project, funded under the European FP7-INFRASTRUCTURES-2012-1 call (GA  No:  312979).  C.  Rodrigues  is  also  part-funded  by Arm  under  a  PhD  Studentship  Agreement.  M. Luj\'an is supported by a Royal Society University Research Fellowship. 
\end{acks}
\bibliographystyle{ACM-Reference-Format}
\bibliography{sample-sigconf.bib}
\end{document}